# Parasocial diffusion: K-pop fandoms help drive COVID-19 public health messaging on social media

Ho-Chun Herbert Chang [a,*], Becky Pham [b], Emilio Ferrara [c]

[a] *Department of Quantitative Social Science, Dartmouth College, Hanover, 03755, NH, USA*
[b] *Annenberg School for Communication and Journalism, University of Southern California, Los Angeles, 90027, CA, USA*
[c] *Thomas Lord Department of Computer Science, University of Southern California, Los Angeles, 90027, CA, USA*

ABSTRACT

We examine an unexpected but significant source of positive public health messaging during the COVID-19 pandemic—K-pop fandoms. Leveraging more than 7 million tweets related to mask-wearing and K-pop between March 2020 and December 2021, we analyzed the online spread of the hashtag #WearAMask and vaccine-related tweets amid anti-mask sentiments and public health misinformation. Analyses reveal the South Korean boyband BTS as one of the most significant driver of health discourse. Tweets from health agencies and prominent figures that mentioned K-pop generate 111 times more online responses compared to tweets that did not. These tweets also elicited strong responses from South America, Southeast Asia, and interior States—areas often neglected by mainstream social media campaigns. Network and temporal analysis show increased use from right-leaning elites over time. Mechanistically, strong-levels of parasocial engagement and connectedness allow sustained activism in the community. Our results suggest that public health institutions may leverage pre-existing audience markets to synergistically diffuse and target under-served communities both domestically and globally, especially during health crises.

## 1. Introduction

On August 21, 2020, the president of the WHO Dr. Tedros Adhanon tweeted an unprecendented message in regards to public health communication. “Thank you”, he wrote, “#BTS for the uplifting #BTS_Dynamite and for reminding the #BTSARMY and the rest of us to #WearAMask and take care of our health... during this #COVID19 pandemic”. The world’s most popular boyband had just released their newest single “Dynamite”, and in their announcement directly supported mask-wearing as a public health practice. As our study will show, his tweet grew to become the most shared tweet related to mask-wearing.[1]

In times of public health crises, effective messaging of public health practices is essential to combating diseases. Typically, institutional leaders are responsible for communicating best practices. However, the spread of these practices has faced major challenges, such as from misinformation [1]. In the case of mask wearing, anti-maskers that emerged in the early stages of the COVID-19 pandemic has prevented effective adoption of mask wearing in the US, despite documented improvements in prevention outcomes [2,3].

Alarmingly, misinformation in many ways disproportionally impacts the Global South. Even before COVID-19, reports had documented how misinformation causes more harm in aggregation, such as misinformation on WhatsApp leading to mob killings in India [4]. Another case-in-point, Brazil’s public health had been severely impacted by its president supporting claims of hydrochloroquine as a miracle cure if vaporized. Brazil was amongst the highest countries in false claims (third to the USA and India), and growing more precarious as their strands of misinformation were decoupled from other countries [5]. From the individual level, misinformation can be stymied by increased deliberation [6], reliance on reason over emotion [7], and simply thinking about accuracy before resharing [8]. Practically, external factors such as the timing of the information and trustworthiness of the source can impact one’s belief in misinformation [9], along with the channels where misinformation arises from [10]. Recent reports have shown messaging aligned with political [11] and religious [12] values also decrease resistance toward new health practices. In particular, affective political polarization—polarization driven by the dislike of the opposing political party has been shown to modulate beliefs during COVID-19 and behaviors such as vaccination [13]. On social media, information spread can be distorted by asymmetric reactions to opposing political elites [14] and social bots [15]. In contrast, information from third-parties have been found to generate more trust in how Americans received COVID-19-related information, where COVID-19 data from Johns Hopkins university received more trust than the CDC [16]. The importance of third-parties with no political impetus or alignment but with highly mobilized networks and large online traction could be a key for effective dissemination of health information.

* Corresponding author.
*E-mail address:* herbert@dartmouth.edu (H.-C.H. Chang).

[1] https://twitter.com/drtedros/status/1296926289648025601

In parallel, K-pop fandom has been increasingly exerting their powerful influence on social causes internationally. Through their strategic support for artists and enthusiastic online presence, K-pop fans have attracted significant attention from the press recently for their political activism and coordinated collective action, such as their attempt to co-opt former President Trump's 2020 rally in Oklahoma [17], and their 2020 successful raising of over one million USD to support Black Lives Matter (BLM) through the virtual #MatchAMillion campaign [18]. More generally, with the rise of the Internet's power in diffusing information, entertainment content and uniting like-minded people online, research on fandom, fan activism and their powerful impact on social change through digital media started to significantly expand [19,20]. For instance, the 2020 BLM Movement was found to have been facilitated through entertainment accounts on Instagram [21]. Broadly speaking, traditional fandom studies investigates the relationships that fans as audiences have with and around the media and popular culture [22]. Two notable and pertinent threads of investigation are arguably suggested by this body of literature. First, how is digitally based fan activism different or similar to more traditional forms of social movements? Second, how can we assess societal impacts of fan activism?

Past instances where fandoms engage in civic and political action had already been noted earlier in the 1980s for fundraising cause [19], but had not received concomitant research and press attention, probably because these mobilizing efforts lacked or had not utilized digital media in ways that make themselves more visible. Due to the proliferation of digital media and the increasingly blurring boundaries between cultural and political concerns, the meanings of fan activism have been transformed by fans and the publics to be at the intersection of cultural and political participation [19]. Through existing fan structures and fan practices (such as remixing online content, networking based on shared interests) and newer forms of communication (such as social media, spreadable videos and memes), fandoms mobilized themselves – intentionally and creatively – around real-world civic or political issues through metaphors drawn from popular culture. In other words, fandom engagement with commercial popular culture does not necessarily compromise their quality of civic participation [19,23,24]. They are effective in part due to the affective dimension of these campaigns, as fandoms coalesce around an object of desire, and the role of emotion-based storytelling has been documented in health communication, including the COVID-19 vaccines [25].

While the term "political participation" was traditionally linked to governmental institutions and electoral processes, younger generations have creatively become civically and politically engaged in their own ways. Mediated by their personal interests and social networks, they push real change through advocacy and activist campaigns [19]. Notable examples of fan activism for real world change includes the Harry Potter Alliance, the US-based non-profit organization that leverage Harry Potter fan communities to run activist campaigns through paid and volunteer staff since the early 2010s, in protest against the Israeli military in 2010 [19]. In this regard for youth, activism which has shifted from older and more traditional, often partisan, approaches to consensus-based approaches on digital channels. [24]. Thus, another critical concept is that of parasocial ties [26]. Parasocial ties refer to the one-sided relationships that individuals develop with media personalities or fictional characters. These connections are formed through the consumption of media. The term "parasocial" indicates that the relationship is asymmetric, with one party (the viewer or reader) feeling a sense of connection and familiarity, while the other party (the media figure or character) remains unaware of the individual's existence. When dealing with fan identity and fandom engagement, we are also more broadly seeking to understand the role of parasocial relationships in the process of diffusion. The driving question is not just whether fandoms impact society and offer public spaces for political engagement, but how this happens organizationally. A good deal of qualitative research has been devoted toward this. First and foremost, it is shared, specialized activities, in relation to cultural phenomena, that characterize how fandom communities are formed and sustained [27]. These ties form the basis of how fans develop a collective identity, before their networks are mobilized for social change. Both the scope and the life span of these activism campaigns, of course, must be taken into account for their effectiveness assessment [19].

In lieu with a recent call to synthesize big data studies with qualitative research [28], this paper studies the effectiveness of mobilizing K-pop fanbase to promote positive health practices on a global scale. Utilizing the largest public Twitter database on COVID-19, we found that the most important player in public health messaging, specifically mask wearing, was the Korean boyband BTS. We show how Twitter-based messaging targets populations that are previously harder to reach—particularly the Global South. We thus argue that the synergy of public health officials and collective engagement through entertainment media can be an effective way to not just disseminate critical health messages on a large scale, but also to the most vulnerable populations.

## 2. Methods

### 2.1. Dataset

We culled our data from the largest Public COVID-19 Twitter dataset [29], which was constructed using the Twitter streaming API, through a selection of keywords. We further refined the dataset into three subsets, specific to our study.

- Tweets containing the hashtag #WearAMask.
- Tweets from and retweets of important health institutions or figures. These include Dr. Tedros, the WHO, the CDC, and the Twitter accounts of all state-level office for the study of our US-Centric portion.
- Tweets containing K-Pop: Given the large amount of fandom activity on Twitter, well-established hashtags have been used to express these fandoms. We first subset a list of these hashtags (such as #BTS and #BTSArmy), then extracted the top co-occurring words related to these groups. We focused on BTS, BlackPink and Twice, the three most prominent K-pop groups on Twitter, and validated our hashtags were the most popular via snowball sampling.
- Tweets containing references to vaccination (vaccin) and the top vaccine brands (#jnj, #pfizer, #moderna).
- Tweets containing mentions of *Eric Ding* and *Grey's Anatomy*. Using preliminary open-ended data inspection, similar to a grounded-approach, we found these two topics to have generated the most engagement via retweets, among all entertainment references and thus focus on them.

Given that the entire dataset already captures the general COVID-19 related discourse, each of these data subsets intersect with discussion about the pandemic (see Table 1).

### 2.2. Identifying top events

To identify top events, we constructed a time-series based on target hashtags or keywords. For instance, to identify top events, we considered the temporal volume of the hashtag #WearAMask. We binned the hashtag with weekly windows, to identify peaks. For the top 5 weeks, we then extracted the top Tweets to identify what events generated the large volume. We further validate this by shifting the weekly binning by one and two days, to catch viral events that may be missed if the window bisected their initialization of their diffusion. For the top tweets, we ensure all labeled peaks fulfill a 75% threshold, which occurs frequently due to the power law nature of retweeting.

**Table 1**
Top Hashtags within the K-Pop dataset.

| Hashtag | Count | Hashtag | Count |
|---|---|---|---|
| bts | 437,563 | blackpink | 28,731 |
| covid19 | 177,608 | Bangtan Sonyeondan | 26,672 |
| btsarmy | 162,488 | btsongma | 20,828 |
| unga | 74,745 | musicbank | 18,747 |
| wearamask | 64,479 | lifegoeson | 18,092 |
| coronavirus | 59,486 | covid | 17,963 |
| bts_dynamite | 48,872 | kpop | 17,185 |
| stayhome | 38,660 | dynamite | 16,954 |
| mtvhottest | 32,939 | howyoulikethat | 15,062 |
| army | 29,145 | twice | 14,716 |

### 2.3. Geo-location

Depending on the data subset, only 3.0%–5.2% of tweets have geo-location metadata. We augment location information using a geotagging heuristic (weighing longitude and latitude data, the Google Maps API, then a dictionary-based approach based on the user-provided location [30]). This yields 31.4% country-level tags and 5.9% State-level tags, specific to the USA.

### 2.4. Assessing inequality

To assess K-Pop's impact globally, we take a straightforward approach—by comparing the representation of different countries when K-Pop is include and when they are not. We take the top 14 countries in volume that include K-Pop, which include the United States, Indonesia, India, the Philippines, Brazil, South Korea, Mexico, Peru, Malaysia, Argentina, UK, Thailand, Vietnam, and Colombia. We then perform cross-sectional analysis across these countries.

### 2.5. Political labeling

Here, we elaborate our calculations for the political diet of users using the political leanings of media sources provided by third-party service Media-Bias/Fact-Check. The website compiles a list of media sources classified under left (−2), left-center (−1), center (0), right-center (1), and right biased (2). Sources that are left and right biased are "moderately to strongly biased", and may publish "misleading reports... that damage their [political] cause" and objectives. Left-center and right-center sources on the other hand have "slight to moderate" bias and are "general trust-worthy". Sources tagged as center are generally trustworthy and contain the least bias. Since we are interested in ideological predilection, rather than the trustworthiness of sources, political diet aggregated over a sufficiently long time can serve as a proxy [31]. To calculate a user-level statistic, we utilize the entire dataset of more than 4 billion tweets, then generate a weighed score for each user. Let $U(i)$ be the URLs that user $i$ retweeted, and denote $v(j) \in [-2,-1,0,1,2]$ as the political position of URL $j$. Then, the weighted political position $p$ is given in Eq. (1) as:

$$p(i) = \frac{1}{|U(i)|} \sum_{j \in U(i)} v(j) \quad (1)$$

We then normalize this score between −1 and 1, then round to the nearest integer (−1, 0, 1). We then end up with three labels, color coded with left leaning (−1) as blue, center as pink (0), and right-leaning as red (1). This and similar approaches to weighted ideological scores has been employed in similar studies [30,32–34], and to validate, three coders have gone through 200 tweets each, and found an average accuracy of %82.

## 3. Results

### 3.1. Temporal dynamics

Fig. 1(a) shows the time-series of #WearAMask and the top events associated with it. Top events were extracted by examining the top retweets, such that they fulfill a 75% threshold, which occurs frequently during these spikes due to the power-law distributed nature of retweets and subsequent cumulative proportions. Four events and their time periods stand out. First, near early July, US comedian Randy Rainbow released a parody of the song Raindrom, that both promoted mask wearing. Importantly, this is the key event, in conjunction with other celebrities such as the Grey's Anatomy star Patrick Dempsey and Grammy Award winning singer John Legend that propelled the term into more widespread use.

The third event is the aforementioned tweeting of Dynamite by Dr. Tedros, during mid-August. It is the fourth most significant event related to mask wearing, and shortly afterward, BTS spoke at the United Nations General Assembly and became the most significant event related to mask wearing. This suggests the combination of health officials can leverage the entertainment industry. Similarly, the fourth most significant event (second by volume) was when Dr. Eric Ding retweeted comedy show *Saturday Night Live*'s parody of a Disney song to promote mask wearing. By then, Ding had already attracted a large following regarding COVID-19, due to his accurate commentary and predictions about the pandemic. We investigate the synergy further in the next section.

Fig. 1(b) further shows how BTS performs against other popular K-Pop fandoms on Twitter. BTS has 35.1 million users, BlackPink 6.6 million, and Twice 7.8 million. If we expect the activity from each group to be proportional to their followers, then we would expect a distribution ratio of 71%, 13% and 16%. However, we observe BTS occupying 92% of COVID-19 related discourse. Other ways for which BTS and their fans impacted COVID-19 was by donating to COVID-19 relief in early March, and matched donations of fans to the Black Lives Matter protests in May 2020 and the Yemen Crisis in collaboration with UNICEF. It is important to note that BTS's official account never made an explicit tweet about mask wearing. Rather, these tweets were driven solely by health official's inclusion of a BTS's speeches and the subsequent reaction from their fans. Especially with the inclusion of online donations, the grassroot nature of the movement suggests that beyond messaging users of the fandom were actively taking action.

A natural question that emerges is how the audience of BTS diverge from the other groups shown in Fig. 1. In Fig. 2, we further investigate the organizational network structure of the audiences following BTS (red), Dr. Eric Ding (blue), and Grey's Anatomy (yellow). Dr. Eric Ding is a public health scientist that rose to prominence for his pandemic-related commentary, due to his role as the Chief of the COVID Response Task Force at the New England Institute. In early 2020, he was one of the first to point out the high reproduction index ($R_0 = 3.8$) and compared it to the 1918 influenza, which prompted criticism from epidemiologists for being alarmist. When his statements proved accurate, he arose to popular status especially as a proponent of preventative policy measures. From a network perspective, this increased his level of embeddedness through followers, and a subsequent "audience [1,35]". However, even across audiences, the structure and composition organizationally may diverse.

Fig. 2(a) shows the retweet network, where nodes are users and ties are retweets. For the sake of visualization, the order of coloring is BTS (red), Eric Ding (blue), and Grey's Anatomy (yellow). The audiences are clearly clustered, which suggest greater organizational complexity. For instance, BTS's network structure features more centers of diffusion as compared to both Eric Ding and for Grey's Anatomy, who only have singular centers. This structure is quantitatively validated in 2(b), which describe nodes who have at least $k$ interactions. K-pop in this case dominates the proportion of the network, despite having less representation when $k = 1$ as compared to Dr. Ding. This indicates while those who pay attention to Dr. Ding tweet in high volume, his audience lacks organizational structure – and hence a community – compared to BTS. In this regard, the underlying fandom becomes a crucial organizational support. We characterize their fandom structure further in Section 3.5.

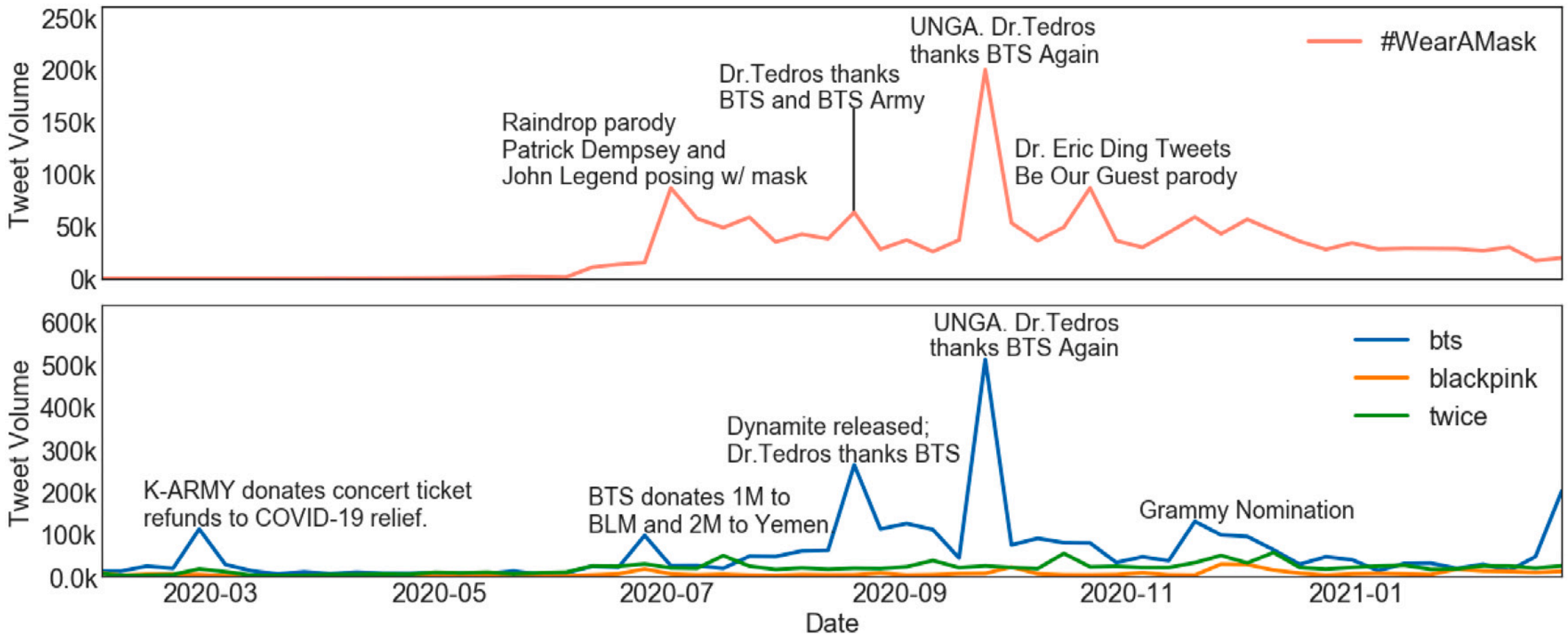

**Fig. 1.** The time series of key terms between March 2020 and March 2021. Fig. 1(a) shows the evolution of the hashtag #wearamask, annotated with the key events during the evolution. Fig. 1(b) shows the time series of top K-Pop groups, annotated with the key events for BTS.

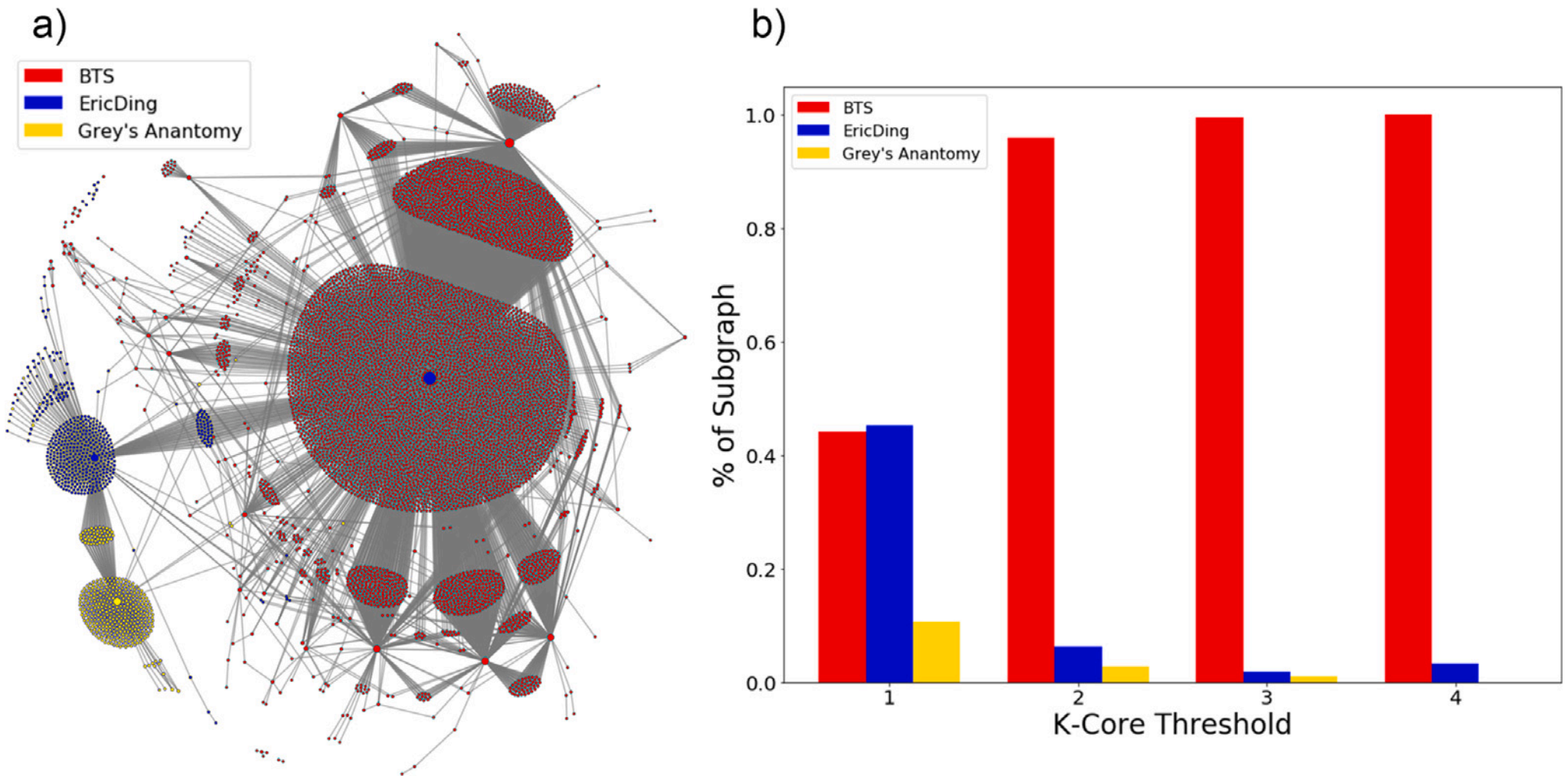

**Fig. 2.** (a) The retweet network following three topical audiences that intersect with #WearAMask—BTS, Eric Ding, and Grey's Anatomy. (b) The community proportions after K-Core decomposition. The size of retweeters of BTS occupy a more central role in terms of information facilitation. (For interpretation of the references to color in this figure legend, the reader is referred to the web version of this article.)

### 3.2. Quantifying synergy

In the time-series, a commonality between the top social media events is the presence of health institutions and the entertainment industry. Whether it is Dr. Tedros thanking BTS or Dr. Eric Ding retweeting *Saturday Night Live*, the combination of entertainers and health officials generate synergistic virality. This motivates the quantification of synergy derived from entertainment. To assess the direct impact BTS had on public health messaging, we consider a group of 70 health agencies and officials based in the United States. We use the United States mainly as the country includes the largest amount of Twitter users in the aggregate data set, and would be at the minimum representative of the impact on a single country. Fig. 1(c) shows the comparison between the retweet volume of tweets including BTS and those that do not. $S(A+B)$ denotes the synergy of $A$ referencing $B$ in a tweet, formally:

**Table 2**
Synergy Calculation of Tweets that include references to K-Pop, primarily BTS.

| | Tweets by health officials | |
|---|---|---|
| | K-pop (B) | No K-pop ($\neg$ B) |
| Total RT Volume | 234,601 | 282,650 |
| Unique Tweets | 16 | 2144 |
| Tweet per post | 14,662.56 | 131.83 |
| $S(A+B)$ | 111.22 | |

$$S(A+B) = \frac{RT(A \cup B)}{RT(A \cup \neg B)} \cdot \frac{T(A \cup \neg B)}{T(A \cup B)}$$

where $RT$ specifies the number of retweets and $T$ the number of unique tweets from $A$. This captures the ratio of average retweets per tweet, conditional on including K-Pop or not (see Table 2).

The first column shows that a total of 16 tweets containing BTS is responsible for 234,601 retweets. On the other hand, there were 2144 tweets during this one-year period that did not feature BTS, which amounted to 282,650 retweets. We then take the ratio of retweets per post, between those that contain K-Pop and those that do not, which yields an increased virality factor of 111.22. While certainly this is contingent on specific events – such as the release of an album – there is considerable evidence here that demonstrates the use of entertainment platforms can boost public awareness of health practices.

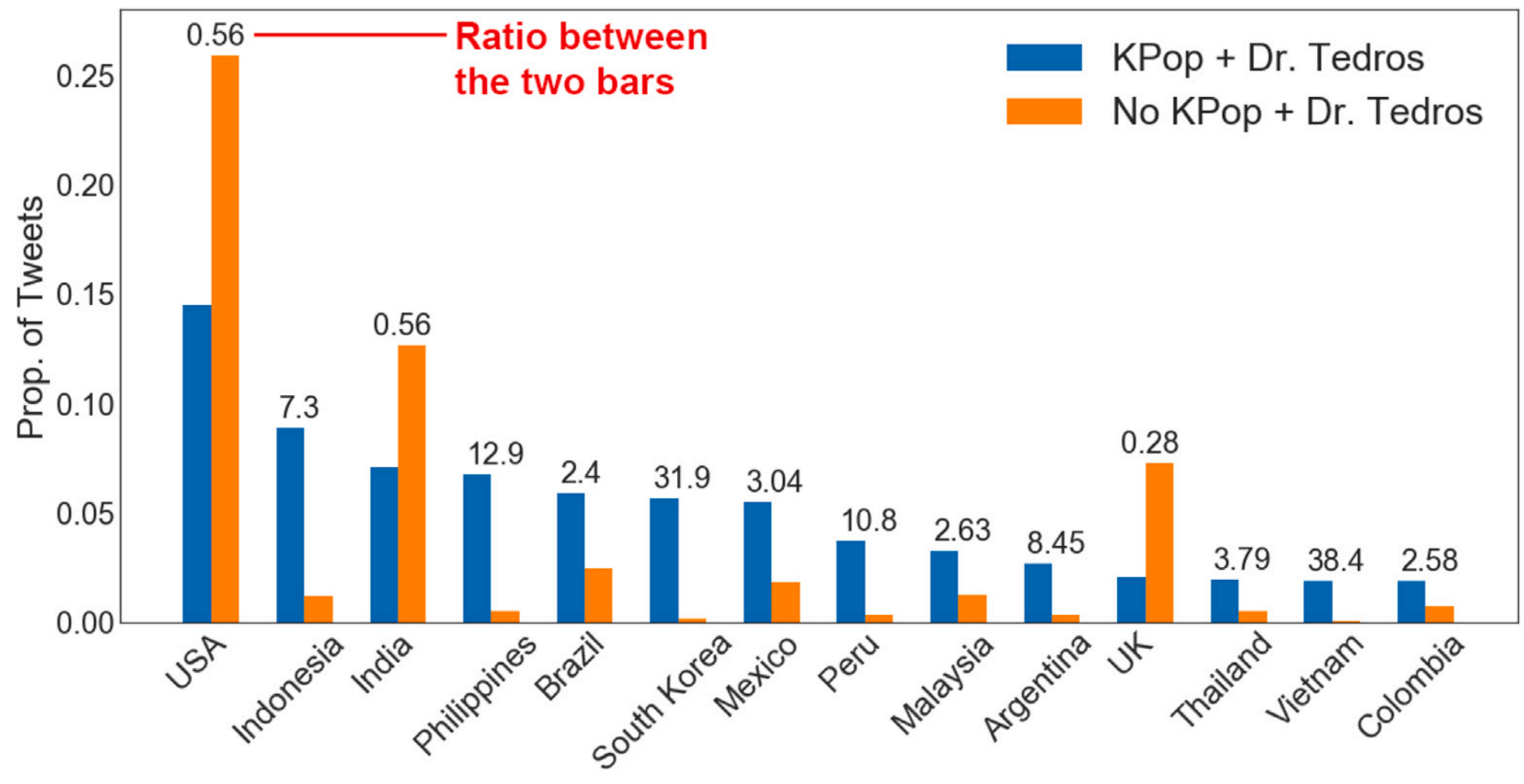

**Fig. 3.** The normalized population size of 14 countries (the union of the top 10 for both categories).

**Table 3**
Percentage advantage of K-Pop-containing tweets. Table 3(a) shows the top 8 states by the population advantage and the associated Twitter dataset advantage. Table 3(b) shows the top 8 States by advantage conferred controlling for the underlying platform (Twitter).

| (a) Top states by diffusion advantage | | | (b) Top states by $\Delta ADV$ | |
|---|---|---|---|---|
| State | $ADV_P$ | $ADV_T$ | State | $ADV_T - ADV_P$ |
| DC | 10.436 | 0.926 | SD | 0.519 |
| NY | 2.058 | 1.323 | ND | 0.410 |
| HI | 1.732 | 1.664 | MS | 0.392 |
| CA | 1.634 | 1.242 | MO | 0.388 |
| ME | 1.306 | 1.219 | UT | 0.374 |
| OR | 1.290 | 0.909 | LA | 0.369 |
| IL | 1.267 | 1.234 | WI | 0.361 |
| TX | 1.209 | 1.086 | NE | 0.329 |

More broadly, this outlines a new framework to understand synergistic diffusion, leveraging advances in machine learning natural language processing. For instance, assuming we can capture policy stances, we can measure different classes of co-diffusion, even antagonistic synergy [36], where the "boost" from adding both is less than the sum. We discuss this further in the conclusion.

### 3.3. Asymmetric diffusion

We have shown BTS increased diffusion dramatically. Next, we investigate geographic localization. Fig. 3(a) shows the normalized population of 14 countries for tweets that include K-Pop and do not include K-Pop. These 14 countries are sorted by those that include K-Pop. While the USA is the top country in both categories, the proportion is much lower for tweets containing K-Pop. We define a ratio that is the percentage with K-Pop over the percentage without K-Pop. We see similar decreases for the UK (0.28) and India (0.56), and dramatic increases in Southeast Asian countries like Indonesia (7.3), the Philippines (12.9), and Vietnam (38.4), and Central/South American countries such as Brazil (2.4), Peru (10.8), and Argentina (8.45).

Table 3 further shows at the State level within the United States, by measuring the proportion of response relative to (a) the population of the United States and (b) the population of the dataset itself. This allows us to separate the influence of the platform from the influence of the content. Moreover, we care about the demographics specific to the platform as it can also be a source of targeted messaging. Using California as an example, let $\%Pop(CA, BTS)$ denote the percentage of BTS tweets arising from California, $\%Pop(CA, T)$ the percentage of all tweets from California, and $\%Pop(CA, P)$ as the total population. Using California as an example, we consider the representational advantage at the state-level.

$$ADV_P = \frac{\%Pop(CA, BTS)}{\%Pop(CA, P)}$$

$$ADV_T = \frac{\%Pop(CA, BTS)}{\%Pop(CA, T)}$$

This first statistic captures the relative representation for messages including K-Pop given the State's population. The second captures the representation relative to the underlying Twitter dataset, which effectively gives the platform-specific expected response. Lastly, by taking the difference:

$$\Delta ADV = ADV_T - ADV_P$$

we effectively remove the representation derived from the underlying platform (Twitter). In other words, if amplification at the population and dataset are equal, then $ADV_T = ADV_P$ or $\Delta ADV = 0$. A more positive value indicates greater bias from the dataset. By observing Table 3a, when comparing BTS and COVID-19 engagement with the State population, we observe much higher diffusion rates in Washington D.C. (1044%), New York (206%), Hawaii (173%), and California (163%), in aggregate.

These advantages may be a result of Twitter being the underlying platform—that is, Twitter itself has a proportionally greater presence in metropolitan areas. When we adjust for the advantage of K-Pop over Twitter ($ADV_T$), we observe that the advantage of $DC$ disappears. Table 3(b) then ranks the top 8 States after controlling for Twitter's underlying advantage. We observe rural interior states, due to the increased rate of diffusion benefit between 132% to 152% compared to their usual rates of response.

This indicates two things. First, the combination of platform and synergistic diffusion can help target States with metropolitan areas for which a greater density of people exist, and thus a greater risk of infection. This is however primarily an effect of using Twitter. Thus, second, the inclusion of BTS also elicits comparatively stronger diffusion into States that are usually neglected by Twitter-based messaging, such as South Dakota, North Dakota, Mississippi, and Utah.

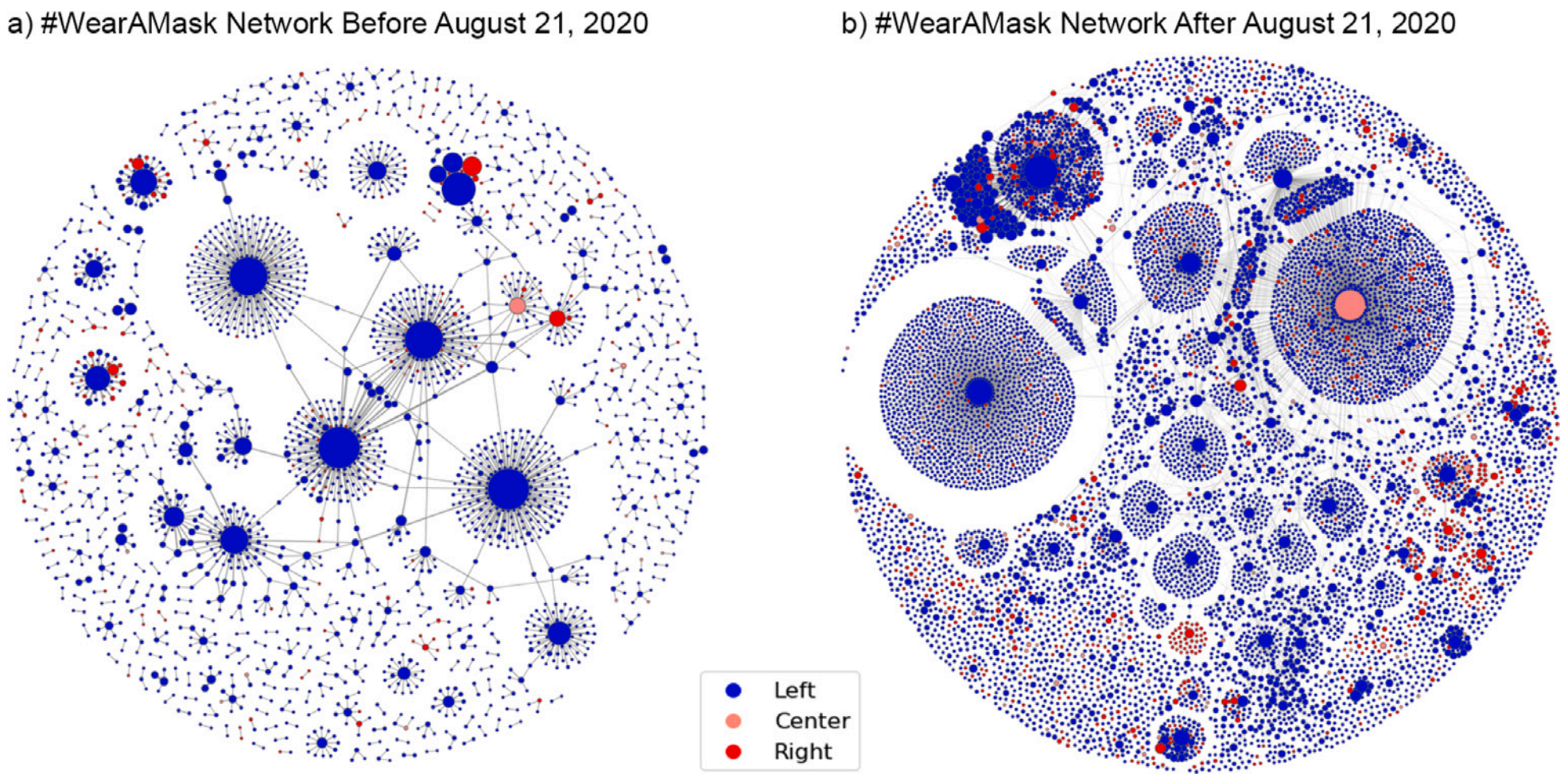

**Fig. 4.** Retweet network of users before (a) and after (b) the tweet by Dr. Tedros within a four-month span. Nodes represent users and are labeled by their political diet, explicitly left-leaning (blue), central (pink), and right-leaning (red). (For interpretation of the references to color in this figure legend, the reader is referred to the web version of this article.)

### 3.4. Shift in ideological alignment and vaccination attitudes

Lastly, we considered the ideological affiliation and network structure of the usage of #WearAMask. Using the retweeting structure of Twitter data, we constructed a network to conduct social network analysis on the users in our misinformation subset. Nodes represent users and links (or edges) represent retweets between users. If user A (tail) retweets user B (head/source), then the strength of their edge is equivalent to the frequency of retweets.

We further calculate the political diet of users using the ideological leanings of media sources provided by third-party service MediaBias/Fact-Check. We end up with three labels, color coded with left leaning (−1) as blue, center as pink (0), and right-leaning as red (1). Pink is chosen as more than 60% of Twitter users are left-leaning, according to the Pew Center. A more detailed description of labeling is given in the methods. Note, ideological leaning is not the same as partisanship (which would be unique to a specific country). The distribution is included in the F 8

Fig. 4 shows the network before and after Dr. Tedros' tweet on August 21, 2020, within a four month window. We use a force-based algorithm through Netwulf [37], a package built on the d3-force module and uses Verlet integration to calculate optimal positions for the n-body problem [38], to plot these networks. Nodes are plotted by their outdegree (how frequently they are retweeted). Immediately, we observe a larger increase in the number of users in the subsequent four months. We also observe strong centers which suggest most users gain their information from select, popular individuals which supports the thesis of Goel, Anderson, Hofman, and Watts [39]. Together, this indicates the reach of the hashtag diffused further.

In both periods, left-leaning users dominate. However, we also observe an increase in the number of red (right-leaning) and pink (center) users. We investigate this further in Fig. 5, plotting the average political scores of normal tweeters (orange) and users who have been retweeted (blue). The $y$-axis indicates political position: the more negative the value, the more left-leaning; the more positive, the more right-leaning.

Whereas the average political position of normal users (orange) who tweet #WearAMask remain largely stable (between −0.9 to 0.77), the political position of users who were retweeted shift drastically from −0.95 in April to −0.65, which is significant given the sample size. If we are to take those retweeted as elite users, given their popularity, this suggests more elite right-leaning users began to significantly use the hashtag more frequently, despite initial reluctance by right-leaning users to take-up on mask-wearing [29]. Additionally, the large upward trend in September comes directly after BTS's appearance on the United Nations General Assembly, though given the many potential confounding events—such as the 2020 United States Presidential Election. However, this increased acceptance by right-leaning elite users to use the hashtag is significant in itself, and deserves further study.

We also conduct a similar time-series analysis for tweets relating to vaccinations, extracting the top events and visualized in Fig. 6(a). Top events were extracted by examining the top retweets, such that they fulfill a 75% threshold. This comparison is of interest as, BTS was invited once more to the United States General Assembly in 2021, and this time commented about COVID-19 vaccinations [40]. Here, they emerged as the fourth most popular event. It is important to put into perspective what the other events were. The most popular event occurred in December 2020, when vaccines began rolling out. Second, the FDA approved Pfizer vaccines. Third, a series of after effects generate anxiety and the popular singer Ariana Grande also posted online imploring people to be vaccinated.

Fig. 6(b) further shows how the weekly sentiment shifts in response to these events. While less robust compared to hand-coding for vaccine policy stance, this method allows us to capture at the macro-scale how sentiment toward the vaccine shifts. We find in cases like Pfizer's announcement of a viable vaccine (December 2020) yielded an increase in both pro-vaccination and anti-vaccination attitudes. We find a transient increase in Pro-vaccination following Ariana Grande and BTS's appearance in the United Nations General Assembly. In general, while it is less clear than the movement-specific analysis in mask wearing, increased positive sentiment without an associated increase in negative sentiment indicates less polarization toward vaccination as a topic.

### 3.5. Community structure and loyalty as a mechanism

Our prior have established K-Pop, specifically BTS, in driving asymmetric information diffusion toward the global south. However, per Fig. 4, there remain discrepancies to the behavior of not just the top K-Pop accounts, but also across other entertainment groups. Here, we analyze potential mechanisms that may possibly explain their relative dominance over the COVID-19 discourse, and focus on two possibilities. First, grass-roots, user-driven approach lead to more robust information diffusion. Second, effects of fandom loyalty cause generate convergence to the most popular groups like the Matthew Effect.

Table 4 shows the top 10 users shown in Fig. 4, based on the total number of retweets. Three of the top users are found to be prominent figures in health or organizations—Dr. Tedros, UNICEF, and the chief of UNICEF. Two are related to the music industry—Big Hit Music, the Grammy Museum, and the IFPI. More importantly, of the top ten are three K-Pop fan accounts, which are non-institutional accounts. For instance, Choi_bts2 is a human user that provides translation services within the community, and occupies the top number of retweets in this sample. In studies of social movements, this is quite distinct. Prior work by Neumayer and Rossi [41] has shown during the Blockify protests institutional accounts rose up to the top. In contrast, the presence of three unverified, every day users is suggestive of emergent coordination which characterizes grassroot social movements.

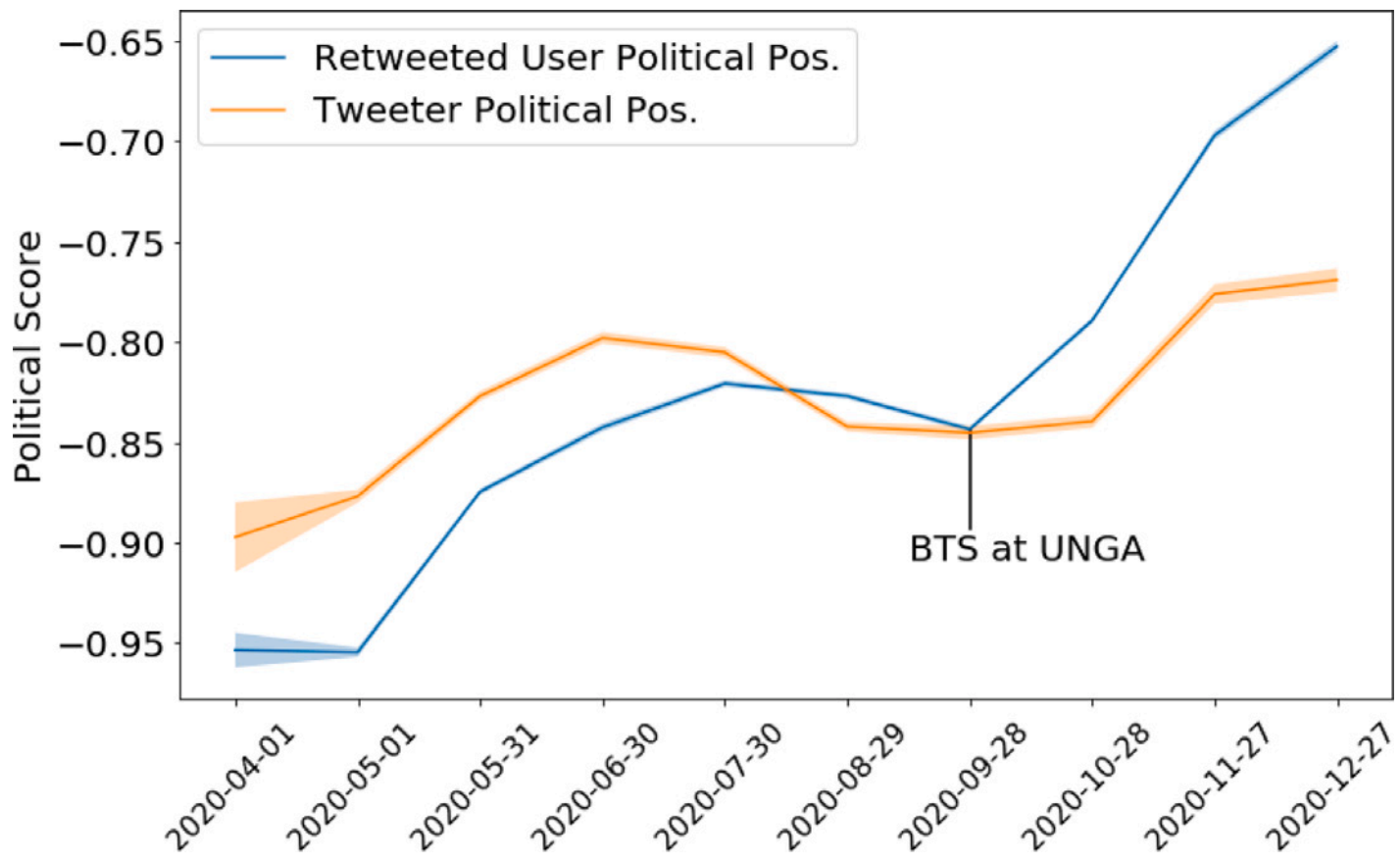

**Fig. 5.** Average political position of normal users (orange) and retweeted users (blue) who use #WearAMask. (For interpretation of the references to color in this figure legend, the reader is referred to the web version of this article.)

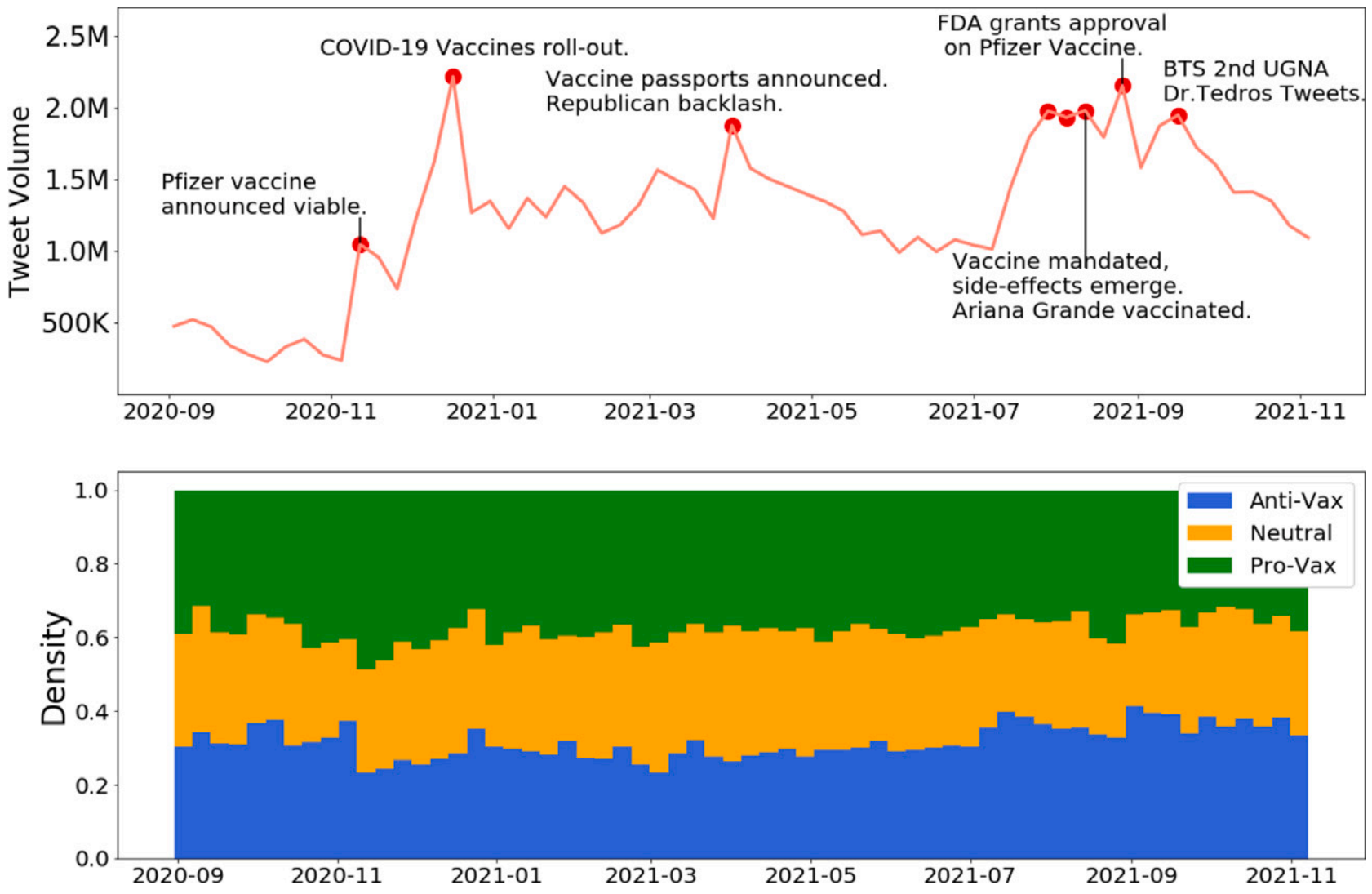

**Fig. 6.** Time-series of vaccination events. (a) shows the top events as related to vaccinations, two months before their the first vaccine was announced viable. (b) shows the sentiment-classified attitudes toward vaccination.

To further characterize fan behavior, we compare the network embeddings of BTS (blue), Black Pink (orange), and Twice (green), the three top K-Pop groups at the time in Fig. 7. Apart from the large mass of BTS followers, what is similar to Fig. 4, we find these communities rarely overlap, with clear clusters. This is particularly noteworthy as, unlike Western-based entertainment groups or entities, all three of these groups come from the same country. It would be reasonable to assume they shared an audience.

We posit this is likely due to the inherent culture surrounding K-Pop fandoms, particularly *stanning* or being a *stan* [42]. A concatenation of “stalker” and “fan”, the term was popularized by rapper Eminem in a song of the same name to describe an overzealous fan. Being a stan demands loyalty toward an entertainment group, which seems to appear macroscopically in this network diagram. The dimension of loyalty is crucial as loyalty is not just a crucial element organizational behavior, but can spur and motivate political action especially in the context of polarization [43–45]. Finding third-party, apolitical communities that demand loyalty may therefore be means of counter-acting issues such as affective polarization, and the resultant resistance toward public health practices.

## 4. Conclusion

The efficacy of public health messaging is vital to responding agilely to health crises, such as the COVID-19 pandemic. In this paper, we investigated the key drivers of the spread and distribution of mask-wearing messaging on Twitter, and if there are any asymmetries in the target audience. Our results indicate the importance of parasocial relationships, and revealed how BTS, a South Korean pop group, became the critical factor for global public health messaging on Twitter. Furthermore, while increased virality is to be expected when influential individuals such as K-pop artists and major Western-based health institutions collaborate, we found the amount of synergy to appear underrated. Importantly, the observed virality in our study does not originate from direct messages from the entertainment groups, but from the community of fans and their grassroots response that suggests potentially significant and tangible results. Although we do not have data on how the K-pop fans in our sample practiced mask-wearing in real life, BTS fandom had previously materialized their online participation from the virtual #MatchAMillion campaign to fundraising of over one million USD for Black Lives Matter [18]. This high-profile demonstration could serve as a benchmark for researchers to gauge and study the possible translation between online collective action and offline results.

**Table 4**
Top users in the K-Pop and COVID-19 network.

| Handle | Description | Total retweets | Total followers | Verified |
|---|---|---|---|---|
| choi_bts2 | A Korean, a BTS ARMY, a fan account, a coffee lover, a human | 350,536 | 1.2M | No |
| UNICEF | United Nations Children's Fund | 185,002 | 8.7M | Yes |
| DrTedros | Director-General for the WHO | 144,858 | 1.4M | Yes |
| unicefchief | Official Account of Henrietta Fore | 129,762 | 85.6K | Yes |
| BigHit_Music | BIGHIT MUSIC Official Twitter. | 98,396 | 20.5M | No |
| allkpop | Breaking kpop celebrity news | 90,327 | 5.7M | Yes |
| GRAMMY-Museum | The Grammy Museum | 58,305 | 75.8K | Yes |
| charts_k | Fan account for BTS. | 57,844 | 599.3K | Yes |
| IFPI_org | Non-profit org. that represents the recording industry | 52,945 | 53.3K | Yes |
| JEONS-GOODE | A personal account, BTS Fan | 49,737 | 131 | No |

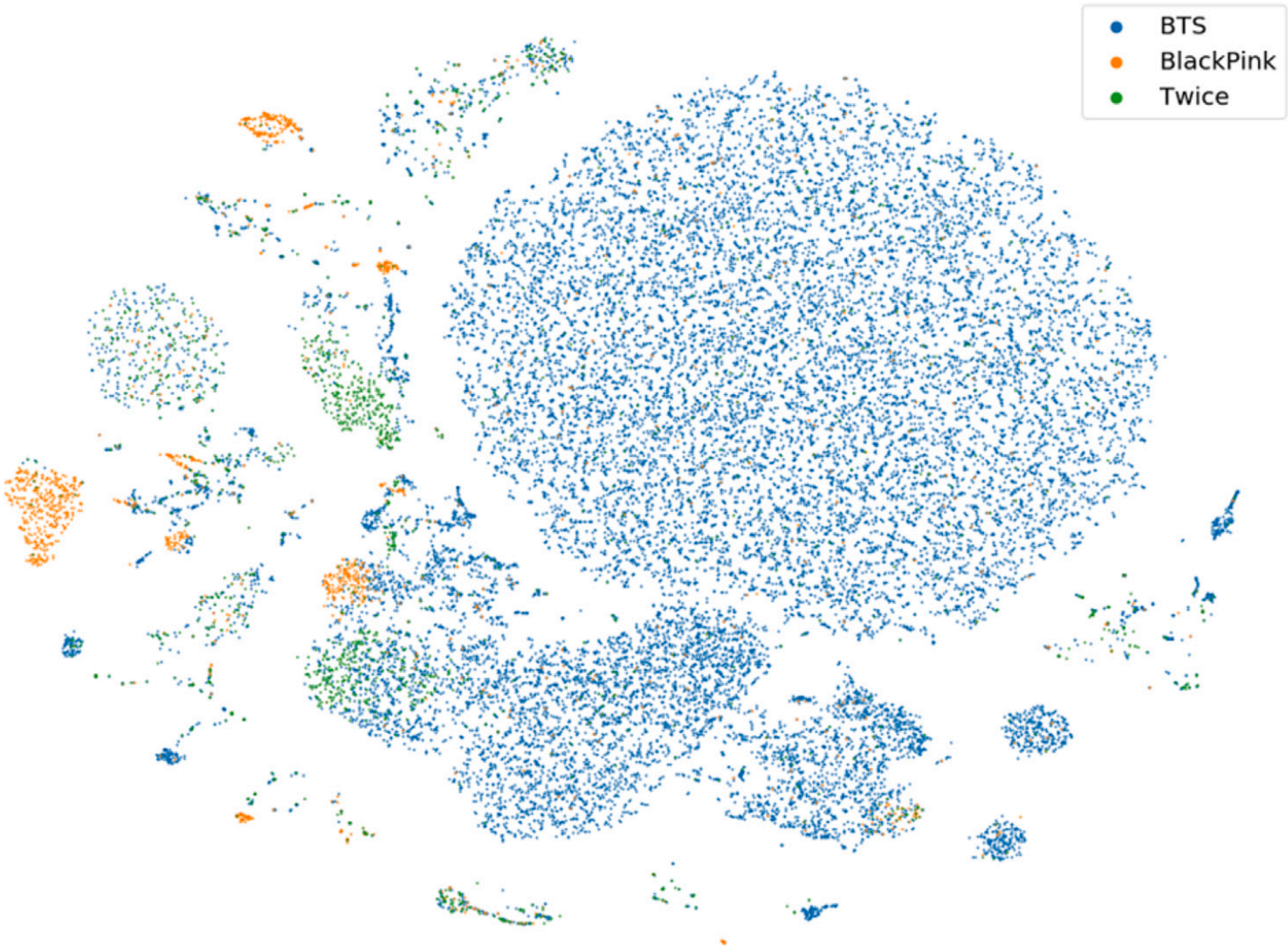

**Fig. 7.** BTS (blue), Black Pink (orange), and Twice (green) network visualization using t-SNE embeddings. (For interpretation of the references to color in this figure legend, the reader is referred to the web version of this article.)

More important than the virality aspect is the noteworthy potential for targeted messaging. We show that messages including K-pop elicited heavy responses from countries in Southeast Asia and South/Central America, as compared to messages that did not. Within the US, we found increased K-pop-related responses in densely metropolitan States that are highly susceptible to disease transmission, and interior, non-coastal states that are often neglected (probably unintentionally) by mainstream social media campaigns and news. In simple terms, the inclusion of K-pop not only increased the depth of diffusion, but also toward more diverse and traditionally under-served areas both globally and domestically.

Preliminary results show the average use of the term shifts toward the political center, especially among right-leaning elite users. This increase coincides with BTS's appearance on the United Nation's General Assembly, though given the number of confounding variables during this period, such as the election. We can conclude the use of entertainment platforms may play a part in depolarization, and deserves further study such as attitudinal shifts through regression discontinuity design. Mechanistically and topologically, the K-Pop community can be characterized by high levels of loyalty, connectivity, and grassroot activity. This positions them, as do other entertainment fandoms or audiences who share these characteristics, as candidates to drive sustained communication in service of public health interventions.

With prior studies showing the importance of perceived neutral third-parties in establishing source trusthworthiness [16], social media use by entertainers, and its spreadability through diverse forms of fan engagement, may be critical to responding to future public health crises. Internationally recognized entertainment artists, evidenced by their role in the COVID-19 pandemic, contain significant broadcasting power due to tightly tuned audiences, intersecting across ethnicity, national identity, political alignment, and beliefs. This area of synergistic communication and diffusion remains an open field of study [36,46,47], and future research entails studying community dynamics more closely, such as their potentially mediating role between health organizations and politicians [48]. The growing ease of using large-language models (LLM) such as BERT [49] and ChatGPT to tackle more nuanced task such as policy stance detection within discourse. This allows more advanced cross-sectional analysis of synergistic diffusion and yield insight in how social media can provide targeted interventions toward the geographic disadvantaged and through political polarization [50,51].

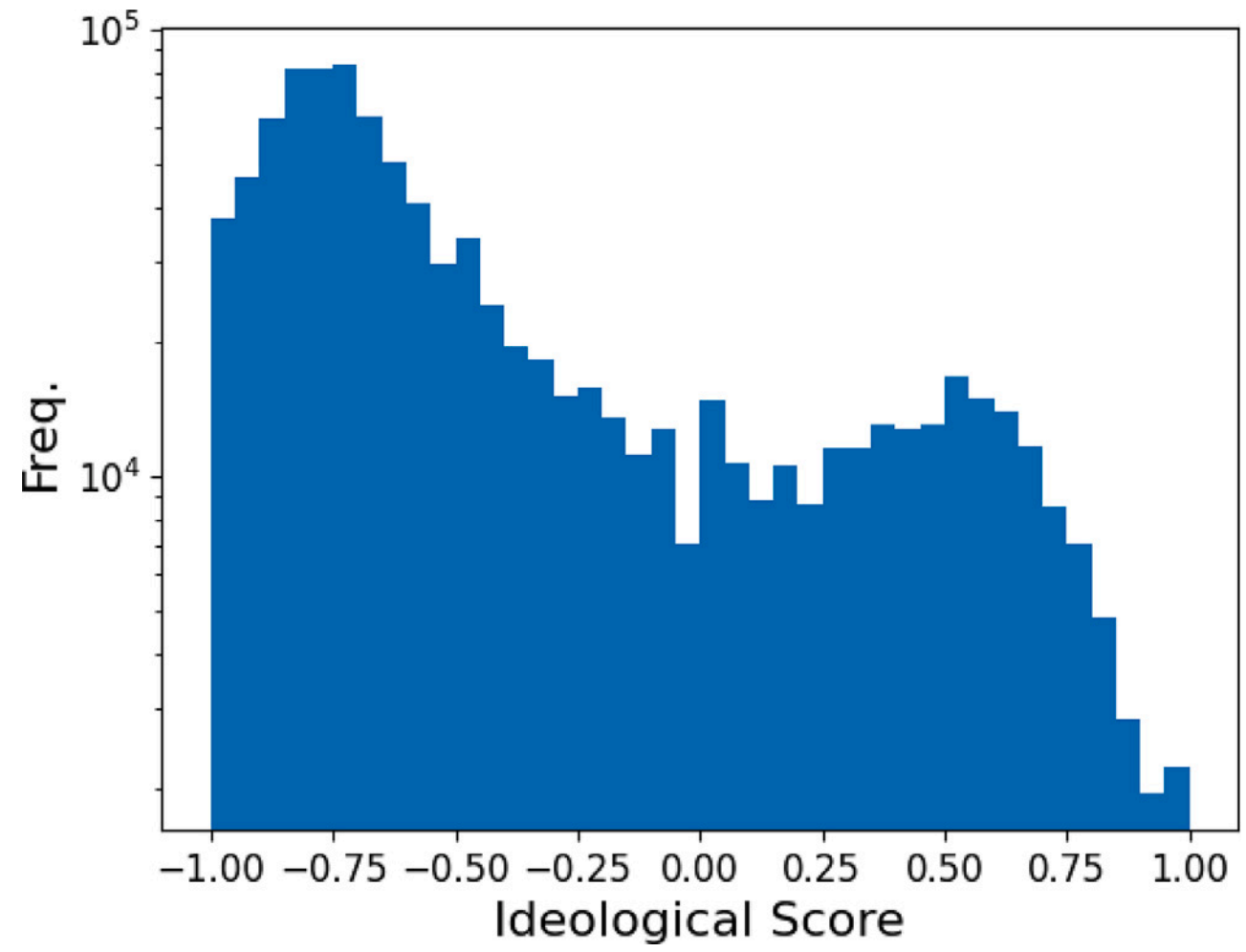

**Fig. 8.** Ideological distribution of users, rated −1 to 1.

## CRediT authorship contribution statement

**Ho-Chun Herbert Chang:** Conceptualization, Data curation, Formal analysis, Investigation, Methodology, Software, Validation, Visualization, Writing – original draft, Writing – review & editing. **Becky Pham:** Conceptualization, Formal analysis, Investigation, Writing – original draft, Writing – review & editing. **Emilio Ferrara:** Conceptualization, Methodology, Writing – original draft, Writing – review & editing.

## Declaration of competing interest

The authors declare that they have no known competing financial interests or personal relationships that could have appeared to influence the work reported in this paper.

## Data availability

I have shared my data and code.

## Acknowledgments

H.C. thanks the Dartmouth Faculty Startup Fund and Walter & Constance Burke Research Initiation Award.

## Appendix

See Fig. 8.